\documentclass[11pt]{article}
\usepackage{graphicx}
\usepackage{amsfonts}
\usepackage{theorem}
\theorembodyfont{\rmfamily}
\newtheorem{Lem}{Lemma}[section]

\newtheorem{The}[Lem]{Theorem}
\newtheorem{Prop}[Lem]{Proposition}

\newtheorem{Rem}[Lem]{Remark}
\newtheorem{Prob}[Lem]{Problem}
\newtheorem{Que}[Lem]{Question}

\begin{document}
\title{On trace inequalities and their applications to noncommutative communication theory}
\author{Kenjiro Yanagi$^1$\footnote{E-mail:yanagi@yamaguchi-u.ac.jp}, Shigeru Furuichi$^2$\footnote{E-mail:furuichi@ed.yama.tus.ac.jp} 
and Ken Kuriyama$^1$\footnote{E-mail:kuriyama@yamaguchi-u.ac.jp}\\
$^1${\small Yamaguchi University, Ube city, Yamaguchi, 755-8611, Japan}\\
$^2${\small Tokyo University of Science, Yamaguchi, 756-0884, Japan}}
\date{}
\maketitle
{\bf Abstract.} Certain trace inequalities related to matrix logarithm are shown.
 These results enable us to give a partial answer of the open problem conjectured by A.S.Holevo.
 That is, concavity of the auxiliary  function which appears in the random coding exponent 
as the lower bound of the quantum reliability function for general quantum states is proven
 in the case of $0\leq s\leq 1$.

\vspace{3mm}
\vspace{3mm}

\vspace{3mm}

%%%%%%%%%%%%%%%%%%%%%%%%%%%%%%%%%%% Section %%%%%%%%%%%%%%%%%%%%%%%%%%%%%%%%%%%%%%%%%%%%%%%%%%%%%%%%%%%

\section{Introduction}
In noncommutative(quantum) communication theory, the concavity of the 
 auxiliary function of the quantum reliability 
function has remained as an open question \cite{Hol2} and unsolved conjecture \cite{ON}.
The auxiliary function $E (s), (0\leq s\leq 1)$ is defined by
\begin{equation}\label{au_func}
E(s) \equiv -\log \left\{\hbox{Tr} \left[\left(  \sum_{i=1}^a \pi_i S_i^{\frac{1}{1+s}}    \right)^{1+s}\right]  \right\},
\end{equation}
where each $S_i$ is the density matrix  and each $\pi_i$ is nonnegative number satisfying $\sum_{i=1}^a\pi_i =1$.
See \cite{Hol3,Hol2} for details on quantum reliability  function theory.
For the above problem,
we gave the sufficient condition on concavity of the auxiliary function 
in the previous paper \cite{FYK}.
\begin{Prop} (\cite{FYK})
If the trace inequality 
\begin{equation} \label{s_con}
\hbox{Tr}\left[A(s)^s \left\{\sum_{j=1}^a 
\pi_jS_j^{\frac{1}{1+s}}\left(\log S_j^{\frac{1}{1+s}}\right)^2\right\}-A(s)^{-1+s}
\left\{\sum_{j=1}^a \pi_j H\left(S_j^{\frac{1}{1+s}} \right)
\right\}^2\right] \geq 0.
\end{equation}
holds for any real number $s \,\,(0\leq s \leq 1)$, any density matrices $S_i (i=1,\cdots ,a)$ 
and any probability distributions $\pi = \left\{\pi_i\right\}_{i=1}^a$, 
 under the assumption that $A(s) \equiv \sum_{i=1}^a \pi_i S_i^{\frac{1}{1+s}} $ is invertible, 
then the auxiliary function $E(s)$ defined by
Eq.(\ref{au_func}) is concave for all $s \,\,(0\leq s \leq 1)$. Where $H(x) = -x\log x$ is the matrix entropy 
introduced in \cite{NU}. \vspace{3mm} \\ 
\end{Prop}

We note that our assumption \lq\lq $A(s)$ is invertible" is not so special condition, because $A(s)$
becomes invertible if we have one invertible $S_i$ at least. Moreover, we have the possibility 
such that $A(s)$ becomes invertible even if all $S_i$ is not invertible for all $\pi_i \neq 0$.
 
In the present paper, we show some trace inequalities related to matrix logarithm,
and then give a partial solution of the open problem in noncommutative communication theory 
as an application of them.

%%%%%%%%%%%%%%%%%%%%%%%%%%%%%%%%%%% Section %%%%%%%%%%%%%%%%%%%%%%%%%%%%%%%%%%%%%%%%%%%%%%%%%%%%%%%%%%%

\section{Main results}
In the previous section, we found that in order to prove the concavity of the 
auxiliary function Eq.(\ref{au_func}), we have only to prove the sufficient condtion Eq.(\ref{s_con})
for any $a$, $s, (0\leq s \leq 1)$ and any density matrices $S_i$.
For this purpose, we consider the simple case $a=2$ and then we put 
$A=S_1^{\frac{1}{1+s}}$, $B=S_2^{\frac{1}{1+s}}$ and $\pi_1=\pi_2=\frac{1}{2}$
 for simplicity. Thus our problem can be deformed as follows:

\begin{Prob}
Prove 
\begin{equation}\label{new_prob}
\hbox{Tr}[(A+B)^s\left\{A(\log A)^2+B(\log B)^2\right\}-(A+B)^{-1+s}(A \log A+B \log B)^2] \geq 0
\end{equation}
for any $s, (0\leq s \leq 1)$ and two positive matrices $A\leq I$ and $B\leq I$.
\end{Prob}

\begin{The}
For two positive matrices $A\leq I$ and $B\leq I$,
Eq.(\ref{new_prob}) holds in the case of $s = 1$:
$$ %\begin{equation}\label{eq_s1}
\hbox{Tr}[(A+B)\left\{A(\log A)^2+B(\log B)^2\right\}-(A\log A+B \log B)^2] \geq 0.
$$ %\end{equation}
\label{th:theorem1}
\end{The}

\begin{flushleft}
{\bf Proof of Theorem \ref{th:theorem1}.}  
Eq.(\ref{new_prob}) can be directly calculated by
\begin{eqnarray}
&   & \hbox{Tr}[(A+B)^s\left\{A(\log A)^2+B(\log B)^2\right\}] -\hbox{Tr}[(A+B)^{-1+s}(A\log A+B\log B)^2]  \nonumber \\
& = & \hbox{Tr}[(A+B)^{-1+s}(A+B)\left\{A(\log A)^2+B(\log B)^2\right\}] \nonumber \\
&   &  -\hbox{Tr}[(A+B)^{-1+s}(A\log A+B\log B)^2] \nonumber \\
& = & \hbox{Tr}[(A+B)^{-1+s}\{ A^2(\log A)^2+AB(\log B)^2+BA(\log A)^2+B^2(\log B)^2 \}] \nonumber \\
&   &  -\hbox{Tr}[(A+B)^{-1+s} \{A^2(\log A)^2+A\log A B\log B + B\log B A\log A + B^2(\log B)^2 \}] \nonumber \\
& = & \hbox{Tr}[(A+B)^{-1+s} \{AB(\log B)^2+BA(\log A)^2 \}] \nonumber \\
&   &  -\hbox{Tr}[(A+B)^{-1+s}A\log A B\log B] - \hbox{Tr}[(A+B)^{-1+s}B\log B A\log A] \nonumber \\
& = & \hbox{Tr}[(A+B)^{-1+s}AB(\log B)^2] + \hbox{Tr}[(A+B)^{-1+s}BA(\log A)^2] \nonumber \\
&   &  -2{\rm Re} \; \hbox{Tr}[A\log A (A+B)^{-1+s} B\log B]. \label{eq:num3}
\end{eqnarray}
Eq.(\ref{eq:num3})  is further calculated for $s = 1$ such as
\end{flushleft}
\begin{eqnarray*}
&   & \hbox{Tr}[AB(\log B)^2]+\hbox{Tr}[BA(\log A)^2]-2{\rm Re} \; \hbox{Tr}[A\log A B\log B] \\
& = & \hbox{Tr}[AB(\log B)^2] + \hbox{Tr}[BA(\log A)^2]-2{\rm Re} \; \hbox{Tr}[B^{1/2}A^{1/2}\log A A^{1/2}B^{1/2}\log B] \\
& \geq & \hbox{Tr}[AB(\log B)^2] + \hbox{Tr}[BA(\log A)^2] -2(\hbox{Tr}[BA(\log A)^2])^{1/2}(\hbox{Tr}[AB(\log B)^2])^{1/2} \\
& = & \{(\hbox{Tr}[BA(\log A)^2])^{1/2}-(\hbox{Tr}[AB(\log B)^2])^{1/2} \}^2 \geq 0. 
\end{eqnarray*}
Cuachy-Schwarz inequality:
$$
\vert \hbox{Tr} \left[ X^* Y \right] \vert^2 \leq  \hbox{Tr} \left[ X^* X \right] \hbox{Tr} \left[ Y^* Y \right] 
$$
for the matrices $X$ and $Y$, has been applied in the above calculation.
 \hfill q.e.d. 

\begin{Rem}\label{re:remark2}
After the manner of {\rm Theorem} \ref{th:theorem1}, we can prove Eq.(\ref{s_con}) in the case of $s=1$ for any density matrices $S_i$ and any probability distributions 
$\pi=\left\{\pi_i\right\}$, $(i=1,2,\cdots ,a)$,
since the left hand side of Eq.(\ref{s_con}) can be directly calculated in the following
$$
\sum_{i<j}\pi_i\pi_j\left\{\hbox{Tr}\left[S_i^{\frac{1}{2}}S_j^{\frac{1}{2}}\left(\log S_j^{\frac{1}{2}}\right)^2\right]+\hbox{Tr}\left[S_j^{\frac{1}{2}}S_i^{\frac{1}{2}}\left(\log S_i^{\frac{1}{2}}\right)^2\right]-2 {\rm Re} \hbox{Tr}\left[S_i^{\frac{1}{2}}\log S_i^{\frac{1}{2}} S_j^{\frac{1}{2}}\log S_j^{\frac{1}{2}}\right]\right\}.
$$
That is, the extended version of {\rm Theorem} \ref{th:theorem1} holds, by applying Cuachy-Schwarz inequality to the third term in the brace of the above, 
after we slightly performed changes as similar as the proof of {\rm Theorem} \ref{th:theorem1}.
\end{Rem}

\begin{The}
For two positive matrices $A\leq I$ and $B\leq I$,
Eq.(\ref{new_prob}) holds in the case of $s = 0$:
$$ %\begin{equation}\label{eq_s0}
\hbox{Tr}[\left\{A(\log A)^2+B(\log B)^2\right\}-(A+B)^{-1}(A\log A+B\log B)^2] \geq 0.
$$ %\end{equation}
\label{th:theorem2}
\end{The}

\noindent
To prove Theorem \ref{th:theorem2}  we require the following lemma.

\begin{Lem} (\cite{And,HP})
For the continuous function $f:[0,\alpha) \to {\bf R}$, $(0 < \alpha \leq \infty)$, 
the following statements are equivalent.
\begin{description}
\item[(i)]  $f$ is operator convex and $f(0) \leq 0$.
\item[(ii)]  For the bounded linear operators $K_i,\,\, (i=1,2,\cdots ,n)$ satisfying 
$\sigma(K_i) \subset [0,\alpha)$,
where $\sigma(Z)$ represents the set of all spectrums of the bounded linear operator $Z$,  
and the bounded linear operators  $C_i,\,\, (i=1,2,\cdots ,n)$  satisfying $\sum_{i=1}^nC_i^*C_i \leq I$, we have
$$
f(\sum_{i=1}^n C_i^*K_iC_i) \leq \sum_{i=1}^n C_i^*f(K_i)C_i .
$$
\end{description}
\label{lem:lemma1}
\end{Lem}

\begin{flushleft}
{\bf Proof of Theorem \ref{th:theorem2}.}  
For $C_1 = A^{1/2}(A+B)^{-1/2}$ and $C_2 = B^{1/2}(A+B)^{-1/2}$,  we have 
$C_1^*C_1 + C_2^*C_2 = I$. Note that $A \leq I$ and $B \leq I$. 
Then we set  $f(t) = t^2, K_1 = -\log A$ and $K_2 = -\log B$ and then apply
Lemma \ref{lem:lemma1}. 
Thus we have
\end{flushleft}
\begin{eqnarray*}
&& \left\{(A+B)^{-1/2} A^{1/2}(-\log A) A^{1/2}(A+B)^{-1/2}+
(A+B)^{-1/2} B^{1/2}(-\log B) B^{1/2}(A+B)^{-1/2}\right\}^2 \\
&& \leq  (A+B)^{-1/2} A^{1/2}(-\log A)^2 A^{1/2}(A+B)^{-1/2}+
(A+B)^{-1/2} B^{1/2}(-\log B)^2 B^{1/2}(A+B)^{-1/2}.
\end{eqnarray*}
Since $\left[A^{1/2}, \log A \right]=0$ and $\left[B^{1/2}, \log B \right]=0$, we have
\begin{eqnarray*}
&& \left\{(A+B)^{-1/2}(-A\log A-B\log B)(A+B)^{-1/2}\right\}^2 \\
&& \leq  (A+B)^{-1/2}\left\{A(-\log A)^2+B(-\log B)^2\right\}(A+B)^{-1/2}.
\end{eqnarray*}
That is,
\begin{eqnarray*}
&& (A+B)^{-1/2}(A\log A+B\log B)(A+B)^{-1}(A\log A+B\log B)(A+B)^{-1/2} \\
&& \leq  (A+B)^{-1/2}\left\{A(\log A)^2+B(\log B)^2\right\}(A+B)^{-1/2}.
\end{eqnarray*}
Thus we have
\begin{equation}\label{opeq_s0}
(A\log A+B\log B)(A+B)^{-1}(A\log A+B\log B) \leq A(\log A)^2+B(\log B)^2. 
\end{equation}
Therefore, if we take the trace in the both sides, then the proof is completed. 
\ \hfill q.e.d. 

\begin{Rem}
After the manner of {\rm Theorem} \ref{th:theorem2}, we can prove Eq.(\ref{s_con}) in the case of $s=0$ 
for any density matrices $S_i$ and any probability distributions $\pi=\left\{ \pi_i\right\}$, $(i=1,2,\cdots ,a)$,
since Lemma \ref{lem:lemma1} is available for any finite number $n$. 
Indeed, we can apply Lemma \ref{lem:lemma1} by putting $K_i = -\log S_i, C_i = \pi_i^{1/2}{S_i}^{1/2}\left(\sum_{k=1}^a\pi_kS_k\right)^{-1/2}$ for $i=1,2,\cdots ,a$ and $f\left(t\right) = t^2$. 
\label{re:remark3}
\end{Rem}

\begin{Que}
From Eq.(\ref{opeq_s0}), the matrix inequality holds in the case of $s=0$. However, 
we do not know whether the following matrix inequalities
\begin{equation}\label{q1}
\left( {A + B} \right)^{1/2} \left\{ {A\left( {\log A} \right)^2  
+ B\left( {\log B} \right)^2 } \right\}\left( {A + B} \right)^{1/2}  
\ge  \left( {A\log A + B\log B} \right)^2  
\end{equation}
or
\begin{equation}\label{q2}
\left\{ {A\left( {\log A} \right)^2  + B\left( {\log B} \right)^2 } \right\}^{1/2} 
\left( {A + B} \right)\left\{ {A\left( {\log A} \right)^2  + B\left( {\log B} \right)^2
 } \right\}^{1/2}  \ge  \left( {A\log A + B\log B} \right)^2 
\end{equation}
corresponding to the case of $s=1$ for any two positive matrices $A\leq I$ and $B\leq I$ hold or not.
We have not yet found any counter-examples,
 namely the examples that the matrix ineqalities both Eq.(\ref{q1}) and Eq.(\ref{q2})
 are not satisfied simultaneously, for some positive matrices $A\leq I$ and $B\leq I$.
\end{Que}

\begin{The}
Suppose $A$ and $B$ are $2\times 2$ positive matrices. Then for any $0 \leq s \leq 1$ we have
$$
\hbox{Tr}[(A+B)^s\left\{A(\log A)^2+B(\log B)^2\right\}-(A+B)^{-1+s}(A\log A+B\log B)^2] \geq 0.
$$
\label{th:theorem3}
\end{The}

\noindent
{\bf Proof of Theorem \ref{th:theorem3}.}  We consider the Schatten decomposition of $A+B$ as follows:
\begin{equation}\label{Sch}
A+B = \sum_n t_n \vert \phi_n\rangle \langle \phi_n\vert ,
\end{equation}
where $\{ t_n \}$ are the eigenvalues of  $A+B$, $\{ \vert \phi_n\rangle  \}$ are the corresponding eigenvectors. 
Then we have
\begin{eqnarray*}
&   & \hbox{Tr}[(A+B)^s\left\{A(\log A)^2+B(\log B)^2\right\}] \\
& = & \sum_n \langle \phi_n\vert (A+B)^{s/2}\left\{A(\log A)^2+B(\log B)^2\right\} (A+B)^{s/2}\vert \phi_n\rangle  \\
& = & \sum_n \langle \phi_n (A+B)^{s/2}\vert \left\{A(\log A)^2+B(\log B)^2\right\}\vert (A+B)^{s/2} \phi_n\rangle  \\
& = & \sum_n t_n^s\langle \phi_n\vert \left\{A(\log A)^2+B(\log B)^2\right\}\vert \phi_n\rangle  \\
& = & \sum_n t_n^s a_n.
\end{eqnarray*}
As similarly, we have
\begin{eqnarray*}
&   & \hbox{Tr}[(A+B)^{-1+s}(A\log A + B\log B)^2] \\
& = & \sum_n t_n^{-1+s}\langle \phi_n\vert (A\log A + B\log B)^2\vert \phi_n\rangle \\
& = & \sum_n t_n^{-1+s} b_n.
\end{eqnarray*}
Where we put $a_n = \langle \phi_n\vert \left\{A(\log A)^2+B(\log B)^2\right\}\vert \phi_n\rangle  $ and
 $b_n= \langle \phi_n\vert (A\log A + B\log B)^2\vert \phi_n\rangle $.
The proof is completed by using the following lemma. 
\ \hfill q.e.d

\begin{Lem}
Suppose the positive numbers $t_1, t_2, a_1, a_2, b_1$ and $b_2$ satisfy the following two conditions. 
\begin{description}
\item[(i)]  $t_1a_1+t_2a_2 \geq b_1+b_2$ 
\item[(ii)]  $a_1+a_2 \geq t_1^{-1}b_1+t_2^{-1}b_2$
\end{description}
Then for any $0 \leq s \leq 1$ we have 
$$
t_1^sa_1+t_2^sa_2 \geq t_1^{-1+s}b_1+t_2^{-1+s}b_2.
$$
\label{lem:lemma2}
\end{Lem}
{\bf Proof of Lemma \ref{lem:lemma2}.}  
It is trivial for $t_1 = t_2$ so that we can suppose $t_1 > t_2$ without loss of generality.
From the condition (i), we then have the following
\begin{eqnarray*}
 t_1^sa_1 + t_2^sa_2 - t_1^{-1+s}b_1 -t_2^{-1+s}b_2 
& = & t_1^sa_1- t_1^{-1+s}b_1+ t_2^sa_2 -t_2^{-1+s}b_2 \\
& = & t_1^{-1+s}(t_1a_1-b_1) + t_2^{-1+s}(t_2a_2-b_2) \\
& \geq & t_1^{-1+s}(b_2-t_2a_2) + t_2^{-1+s}(t_2a_2-b_2) \\
& = &(t_2^{-1+s}-t_1^{-1+s})(t_2a_2-b_2).
\end{eqnarray*}
Since $t_2^{-1+s}-t_1^{-1+s} \geq 0$,
if $t_2a_2-b_2 \geq 0$, then the lemma follows. On the other hand, if $t_2a_2-b_2 < 0$, from the condition (ii) we then have
\begin{eqnarray*}
t_1^sa_1 + t_2^sa_2 - t_1^{-1+s}b_1 -t_2^{-1+s}b_2 
& = & t_1^sa_1- t_1^{-1+s}b_1+ t_2^sa_2 -t_2^{-1+s}b_2 \\
& = & t_1^s(a_1-t_1^{-1}b_1) +t_2^s(a_2-t_2^{-1}b_2) \\
& \geq & t_1^s (t_2^{-1}b_2 -a_2) +t_2^s(a_2-t_2^{-1}b_2) \\
& = & (t_1^s - t_2^s)(t_2^{-1}b_2 -a_2) \geq 0.
\end{eqnarray*}
\hfill q.e.d.

\begin{Rem}
After the manner of  {\rm Theorem} \ref{th:theorem3}, we can prove Eq.(\ref{s_con}) for any $2\times 2$ density matrices $S_i$ and any probability distributions $\pi=\left\{\pi_i\right\}$, $(i=1,2,\cdots ,a)$,
by considering
the Schatten decomposition of the $2\times 2$ positive matrix
$\sum_{k=1}^a\pi_kS_k^{\frac{1}{1+s}}$ as follows: 
$$
\sum_{k=1}^a\pi_kS_k^{\frac{1}{1+s}} = \sum_n \lambda_n|\phi_n\rangle \langle\phi_n|,
$$
where $\lambda_1$ and $\lambda_2$ are the eigenvalues of $\sum_{k=1}^a\pi_kS_k^{\frac{1}{1+s}}$, 
$\{|\phi_1\rangle\}$ and $\{|\phi_2\rangle\}$ are corresponding eigenvectors, respectively.
Therefore it was shown  the concavity of the auxiliary function $E \left(s\right)$ of the quantum reliability 
function for any $2\times 2$ density matrices $S_i$. Thus we gave a partial solution for the open probelm given in \cite{Hol2}.
\end{Rem}
\begin{Rem}
We expect that our {\rm Lemma} \ref{lem:lemma2} can be extended to the general $n \geq 3$,
where $n$ represents the number of the eigenvalues given in Eq.(\ref{Sch}).
However it is impossible to prove it, 
because we have a counter-example for such a generalization.
For example, we take 
$$
s = \frac{1}{2}, t_1=3, t_2=2, t_3=1, a_1=\frac{2}{3}, a_2=1,a_3=\frac{3}{2},
b_1=\frac{1}{2}, b_2=4, b_3=1.
$$
Although it holds two conditions corresponding to the generalization of two conditions 
(i) and (ii) in Lemma \ref{lem:lemma2}:
$$
t_1a_1 + t_2a_2 + t_3a_3 =b_1 +b_2+b_3 = \frac{11}{2}
$$
and
$$
a_1+a_2+a_3 = t_1^{-1}b_1 +t_2^{-1}b_2 +t_3^{-1}b_3=\frac{19}{6} ,
$$
the following calculations:
$$
t_1^s a_1 +t_2^s a_2 +t_3^s a_3=\frac{2\sqrt{3}}{3}+\sqrt{2}+\frac{3}{2} \simeq  4.068914
$$
and
$$
t_1^{-1+s} b_1 +t_2^{-1+s} b_2 +t_3^{-1+s} b_3 = \frac{\sqrt{3}}{6}+2\sqrt{2}+1 \simeq 4.1171021,
$$
show that
$$
t_1^s a_1 +t_2^s a_2 +t_3^s a_3 \geq t_1^{-1+s} b_1 +t_2^{-1+s} b_2 +t_3^{-1+s} b_3
$$
does not hold. This means that our Lemma \ref{lem:lemma2} can not be extended to the general case of $n\geq 3$.
Therefore we must produce an another method to prove Theorem \ref{th:theorem3} for any $n \times n$
positive matrices $A$ and $B$. Our Theorem \ref{th:theorem3} is constructed by a kind of 
the interpolation between two conditions generated by Theorem \ref{th:theorem1} and
Theorem \ref{th:theorem2}. If we extend this method to the case of $n \geq 3$, we may require the further
conditions.
\end{Rem}

\section{The related inequalities}

We introduce the following symbol in the relation to quantum relative entropy.
For the positive matrices $A$ and $B$, we define
$$
D\left(A \| B\right) = A\left(\log A - \log B\right).
$$
Then we have the next theorem.  % due to Theorem  \ref{th:theorem1} and \ref{th:theorem2}. 
\begin{The} 
\begin{description}
\item[(1)]  $\hbox{Tr}[D(A \| B)D(B \| A)] \leq 0$.
\item[(2)]  $\hbox{Tr}[(A+B)^{-1}D(A \| B)D(B \| A)^*] \leq 0$.
\end{description}
\label{th:theorem4}
\end{The}

\begin{Rem}
The quantum relative entropy  is defined by
$H(A \| B) = \hbox{Tr}[D(A \| B)]$ 
for any density matrices $A$ and $B$.
The relative matrix entropy \cite{FK} is defined by 
$$
S(A \| B) = A^{1/2}(\log A^{-1/2}BA^{-1/2})A^{1/2}
$$ 
for any invertible positive matrices $A$ and $B$.
Moreover, if  $A$ and $B$  are commutative, then we have $D(A \| B) = -S(A \| B)$. 
\label{re:remark4}
\end{Rem}

\section*{Acknowledgement}
The authours thank referees for valuable comments on this paper.

\end{document}